\documentclass[utf8, nlm, hcp, author-year, round]{frontiersinFPHY_FAMS} 
\usepackage{url,hyperref,lineno,microtype}
\hypersetup{
    colorlinks=true,
    linkcolor=blue,      
    citecolor=blue,      
    urlcolor=blue        
}
\usepackage[onehalfspacing]{setspace}
\setcitestyle{authoryear,round}

\def\firstAuthorLast{Barentine and Venkatesan}

\def\Authors{John Barentine$^{1,2*}$ and Aparna Venkatesan$^{1,3}$}

\def\Address{$^{1}$Center for Space Environmentalism, Washington, DC, United States \\
$^{2}$Dark Sky Consulting, LLC, Tucson, AZ, United States \\
$^{3}$University of San Francisco, San Francisco, CA, United States}
\def\corrAuthor{} 
\def\corrEmail{}

\address{\Address}
\correspondance{} 
\extraAuth{}

\begin{document}
\onecolumn
\firstpage{1}

\title[The Outer Space Treaty Won't Save Us]{The Outer Space Treaty Won't Save Us From Ourselves}

\author[\firstAuthorLast]{\Authors}

\correspondance{john.barentine@gmail.com}
\address{} 

\correspondance{john.barentine@gmail.com}

\maketitle

\reversemarginpar 
\marginpar{
    \vspace{-13.5cm} 
    \noindent\fontsize{14}{16}\selectfont\textbf{\color{gray} OPINION}
}

\begin{center}
    \vspace{-2em}
    \textit{Accepted for publication in Frontiers in Space Technologies (Space Economy), Vol.~7} \\
    \href{https://doi.org/10.3389/frspt.2026.1748406}{doi: 10.3389/frspt.2026.1748406}
\end{center}
\vspace{1.5em}

\begin{abstract}
    The rapid growth of human activities in outer space sounds urgent alarms around ethical and philosophical issues, particularly concerning space militarization. The present international legal framework governing activities in space, the Outer Space Treaty (OST), views the peaceful exploration of space for scientific research as co-equal to other ‘uses’ entitled to ``due regard'' with respect to ``potentially harmful interference'' on the part of other space actors. The OST is deeply aspirational but has weak enforcement mechanisms, relying at its core on the goodwill of all involved parties as the fundamental basis for accountability. But that framework now faces unsustainable pressures from both public and private interests, and current agreements like the OST may be unable to exert timely, material protections. Terrestrial frameworks of "ethics of deterrence" versus the "ethics of agreements'' are quickly expanding into cosmic environments. We argue for the legal recognition of space as an environment as the basis of any future approach to securing its integrity, and share examples of agreements grounded in peaceful cooperative stewardship of shared environments. These represent potential pathways forward that are ethical and also serve rational self-interest and self-preservation at this crucial juncture for humanity.
\end{abstract}

\section{A Historical Crossroads for Space Exploration}
We are living through a time in which Earth and near-Earth space are facing unprecedented levels of potentially irreversible change. A potent combination of increasing challenges from war, economic instability and displaced populations along with decreasing adherence to longstanding norms of behavior, pacts between nations, and scientific consensus threatens the world order and a baseline of stability and peace for humanity, with conflicts nearly doubling since 2020~\citep{ArmedConflictIndex2025}. We argue here that this trajectory ignores history’s lessons and will lead to failure to meet self-serving military/security goals of nation-states as well as ethical-philosophical principles that guide trust and agreements that endure. In particular, and more pointedly, the outer space governance institutions (and their circuit breakers) developed during the Cold War no longer effectively serve their intended functions.

We are astronomers, not scholars of war or international relations. Yet we are mindful that our science is conducted both in and ``through'' space. The modern history of our discipline is inseparable from the technological developments brought on by those who sought to make a warfighting domain of outer space. Astronomers have at turns been both victims and beneficiaries of this effective status quo. Yet we increasingly recognize that a new era in the human presence in outer space has dawned, and it threatens to sweep aside everything that came before. Under this pressure, the international legal framework that governs outer space appears ill-equipped to bend sufficiently without breaking altogether.

\section{Accelerating Militarization of Space}
Human activity in space has always contemplated warfare. The first human-made objects to leave the Earth's atmosphere were V-2 rockets in the Second World War. While the launch of Sputnik I in October 1957 is mainly remembered as the deployment of the first artificial satellite into Earth orbit, its real achievement was the successful demonstration of Soviet rocket technology that could be used to deliver nuclear weapons over tens of thousands of kilometers. Seventeen years later the Soviet Salyut-3 space station test-fired an aircraft ‘autocannon’, the first (and so far only) space-to-space weapon known to be deployed in orbit. The following decade the Soviet Union planned, but did not successfully test, the Polyus uncrewed weapons platform equipped with a megawatt carbon-dioxide laser and a self-defense cannon. Both the U.S.S.R. and the United States began testing antisatellite (ASAT) weapons in the early 1960s; by the early 2000s, China, India and Israel had also carried out successful tests. And in the 1980s the U.S. famously proposed the Strategic Defense Initiative (SDI) missile-defense program that became popularly known by the nickname ``Star Wars''. 40 years later its ghost returned, rebranded as the ``Golden Dome''.

Previous efforts to ban warfare in space achieved only modest results. The Limited Test Ban Treaty signed in 1963 by the U.S., U.S.S.R. and the United Kingdom effectively ended high-altitude nuclear tests, including  detonations in space. After the U.S. announcement of the development of SDI in March 1983, the ailing Soviet premier Yuri Andropov denounced the move as an ultimately offensive military action and initiated an effort to frame a treaty banning military weapons from space. His successors pushed for adoption of such a treaty through the end of the Soviet Union in 1991. After a series of destructive ASAT tests in the 1990s and 2000s, a United Nations working group adopted a non-binding resolution in 2022 calling for a general prohibition. Opponents downplayed the result in accusing the U.S. and others of limiting meaningful progress toward preventing an arms race in outer space.

\section{What Treaties Address But Did Not Anticipate}
The ``Treaty on Principles Governing the Activities of States in the Exploration and Use of Outer Space, including the Moon and Other Celestial Bodies,'' more commonly known as the ``Outer Space Treaty'' (OST),\footnote{610 U.N.T.S. 205, 18 U.S.T. 2410, 6 I.L.M. 386 (1967).} was negotiated during one of the heated periods of the Cold War. It binds its 118 state parties to promises not to appropriate territory in outer space through claims of sovereignty; not to deploy weapons of mass destruction in space; and to use space ``in the interest of maintaining international peace and security and promoting international co-operation and understanding.'' Yet it was framed primarily by the United States, the Soviet Union and their respective clients and allies in an era when they were the only two launching states in the world. The voices of developing and sovereign Indigenous nations were wholly shut out of the process~\citep{vanEijk2025}. 

Today, roles of the historical superpowers are diminishing globally, with Russia, once an epicenter of advanced science and space exploration, reduced to episodic, unpredictable actions on Earth and in space. Recent chaotic and seemingly arbitrary U.S. decisions impacting science, health, and economies as well as abruptly withdrawing from longstanding treaties and agreements have led to a growing erosion of U.S. standing on the global stage. U.S. and Russian leadership roles in the scientific enterprise may continue to dwindle as China and other nations pull ahead with long-term strategic investment in research initiatives and facilities. The U.S. in particular is lagging significantly in both hard- and soft-power channels, owing to decades of decreasing funding in infrastructure, research and technological development, with dramatic declines in the scientific workforce and funding of key scientific agencies~\citep{Zhang2026}. Simply put, there is little agency left in U.S. agencies, with the risk of, e.g., NASA becoming a mere contracts manager or another paying customer on private rockets.

Furthermore, we live in an era of increasing geopolitical instability. Militant regimes seek nuclear weapons as a form of effective deterrence against hostile foreign attacks. Although speculative at best, it is possible that the major powers are already deploying nuclear weapons in space, in contravention of OST Article IV. Such deployments did happen in the historic past, some of which – such as the U.S. ``Starfish Prime'' nuclear weapon test in 1962 – wreaked havoc on both the space and terrestrial environments. Current conflicts like that between Iran and the West underscore the reasons why regimes seek their own nuclear arsenals: they have a proven deterrent effect on ‘first strike’ attacks. Even nominally defensive actions can threaten the space environment; additionally, the ongoing crowding of low-Earth orbit (LEO) with spacecraft of all kinds, accelerated by rising military uses, could end in catastrophe.

Considered against this backdrop, the OST is ultimately a weak legal instrument. It is not at all clear that it can stand up to new uses of space not anticipated by its framers, including the growing privatization of space exploration. It has neither real enforcement teeth nor a mechanism to definitively resolve disputes among the state parties. Those state parties increasingly look to new agreements outside the OST framework, which may signal the end of the multilateralism that has characterized international space policymaking for the past six decades~\citep{Nelson2020}. Whatever success the OST has achieved is built on behavioral norms that its member states pledge they will follow. To us this calls to mind the League of Nations, which failed to prevent the outbreak of the Second World War. It attempts to conjure an ``ethics of agreements'' that stands counter to an ``ethics of deterrence'' otherwise prevailing among global superpowers. 

A potential new approach offering a transformative path forward is the `ethics of Nature', which addresses the current crisis by acknowledging that traditional Western legal and diplomatic frameworks are insufficient to prevent Earth- and space-based catastrophes. True sustainability requires a fundamental shift: redefining our relationship with outer space through a lens of reciprocity and recognizing the environment’s intrinsic value beyond human exploitation. Without this shift, current frameworks risk fostering adversarial competition for resource extraction and orbital dominance. This approach aligns with the terrestrial 'Rights of Nature' movement, advocating for a governance model that prioritizes environmental integrity over geopolitical or economic gain~\citep{Zartner2024}. This is a founding principle of the recently established Center for Space Environmentalism~\citep{CSE2026}.

As our title reveals, fundamental challenge lies in defending the space environment from failure of human goodwill and trust, rather than defending the space environment from the blind spots of current treaties (whose burgeoning irrelevance arises primarily, barring the outsized role of private interests, from the failure of its opening assumptions of peaceful cooperation). Institutional frameworks can be wholly inadequate yet functionally irreplaceable; in our opinion, the OST remains our best option to date. Any alternative legal framework must address the consequences of rogue actions from bad actors and the failure of goodwill/trust – and accept that these will have mixed results with deterrence and accountability – while also addressing space privatization and making environmental impacts a condition of licensing or approval.

\section{Space Geopolitics and the Way Forward}
The United States accounted for more than 50\% of all LEO launches in 2025, and it operates the clear majority of active payloads currently in orbit. It therefore wields an outsized influence on the space domain — current and future — and its actions demand special scrutiny. This is especially true as the U.S. government cedes ground under its OST Article VI supervisory responsibilities to the demands of private commercial space actors launching from its territory, leaving unclear the channels for regulatory oversight and operational accountability. As experts sound alarm bells about the U.S. military’s lack of modernization and preparedness for contemporary warfare~\citep{Milley2024}, the capacity of the U.S. to advance, much less defend, its scientific, innovation and security interests in space, or meet its self-imposed goals of space supremacy, appear increasingly doubtful. Although U.S.-launched satellites currently dominate LEO, this will change with the arrival of more spacefaring nations invested in sovereignty of infrastructure, data and economic development. But the voices of developing economies new to spaceflight are often drowned out by those of established launching states~\citep{Ibrahim2025,Palit2025}, prompting some to lead by example along their own homegrown paths~\citep{Guesgen2025}.

The Moon and cislunar space are already providing an uneasy preview of future space geopolitics. To date, five nations (the U.S., the former Soviet Union, China, India and Japan) have landed on the Moon, with first-timers India and Japan landing within six months of each other between August 2023 and January 2024. Although we scowl at politically loaded terms like ``space race'' being applied to our companion of over four billion years, we note that at present China is leading in humanity’s planned return to the Moon. For the U.S., whether this arises from a generational lack of investment or from the dominance of one company in government contracts for rocket launches and human spaceflight, the net result is that NASA appears to hold a losing hand and cannot regain momentum without closely examining and learning from its mistakes~\citep{Berger2025}. This falling behind will likely accelerate owing to the language of the OST: the nations that arrive first tend to claim prime real estate for science, resource exploitation and bases, with subsequent nations being relegated to sites and activities that do not interfere with already existing operations. 

Another unsettling glimpse of what future operational cooperation in space (or lack thereof) may look like among competing space nations is found in the dramatic loss of usable airspace since 2022, owing to the rise of global conflicts. Flights around the world have not only jumped in duration and cost; they are increasingly unsafe in some airspaces, leading to last-minute flight plan adaptations by commercial airlines and more countries ceasing air service on some routes~\citep{Yerushalmy2025}. We wonder whether this reflects the future of cislunar and outer space traffic management, with terrestrial conflicts reflected in complex space environments far from home. 

A last arena of potential conflict needing coordinated transparency and accountability is the urgent need to improve tracking and mitigation of space debris. Similar to the lopsided fallout for many nations from climate change owing to the actions of a few, countries that are not space actors — especially in the global South — may be the unwilling final resting place of falling space debris generated by a few Western nations~\citep{Byers2022}.

Closing on a broader, philosophical view: outer space is humanity’s shared heritage, and it is where our shared future lies as well. The OST, although written in the very different world of the mid-1960’s, nevertheless contains timeless principles of peaceful exploration, goodwill and trust among spacefaring nations engaging in scientific activities that benefit all peoples, a point it repeatedly states. Immanuel Kant’s viewpoint is especially relevant here: goodwill reflects our unwavering willingness to act on our deepest convictions and ideals that lie beyond the reach of clever conditions or context-dependent outcomes. 

The resolve and imagination needed to show such goodwill and make agreements in good faith are needed now more than ever, and could draw strength from the many examples of these that are thriving. These include the peaceful management of Antarctica as a global scientific commons; collaborative support among astronauts on the International Space Station regardless of the geopolitical challenges of their countries; and the growing models of partnership between Indigenous communities and research facilities on their ancestral lands~\citep{Cuby2024}.

Despite the OST being modeled on the Antarctic Treaty (1959)~\citep{DeSimone2022} and the examples of global scientific cooperation in these Earth/LEO environments, we recognize that humanity’s exploration of new space environments may require new treaties and cooperative agreements. These provide inspiration for the management and stewardship of environments shared by constituencies who build pathways forward that benefit all involved despite their differences.  Such models are critical for our shared future in the exploration and protection of space as an environment. The story of humanity in space is still being written, requiring innovation and determination from institutions and communities alike.

\section*{Conflict of Interest Statement}
Author JB was employed by Dark Sky Consulting, LLC. The remaining author declares that the research was conducted in the absence of any commercial or financial relationships that could be construed as a potential conflict of interest.

\section*{Author Contributions}
JB: Writing – original draft, Writing – review and editing, Conceptualization. AV: Writing – review and editing, Writing – original draft, Conceptualization.

\section*{Funding}
AV gratefully acknowledges the University of San Francisco Faculty Development Fund, and the Kavli Institute of Theoretical Physics at UC Santa Barbara where this work was begun.

\section*{Acknowledgments}
The authors wish to thank the reviewers for their insightful comments which improved the clarity of this manuscript. 

\bibliographystyle{Frontiers-Harvard}
\bibliography{BarentineVenkatesanFrontiers2026}

\end{document}